\documentclass{article}
\usepackage{spconf,amsmath,graphicx}

\usepackage{enumitem}
\usepackage{xcolor}
\definecolor{forestgreen}{RGB}{0,100,0}
\setlist{nosep, leftmargin=14pt}
\usepackage{graphicx} 
\usepackage{mathtools} 
\usepackage{siunitx}
\usepackage{threeparttable}
\usepackage{cite}
\usepackage{graphicx}
\usepackage{setspace}
\usepackage{acronym}
\usepackage{amsfonts}			
\usepackage{times}		
\usepackage{amssymb}
\usepackage{amsmath}
\usepackage{gensymb}
\usepackage{tabularx} 
\usepackage{multirow}
\usepackage{algorithm}
\usepackage{algorithmic}
\usepackage{booktabs}
\usepackage[T1]{fontenc}

\usepackage{float}
\usepackage{mwe}
\newcolumntype{L}{>{\centering\arraybackslash}m{3cm}}
\usepackage[thinc]{esdiff}
\usepackage[table]{xcolor}
\usepackage{soul}
\usepackage{textcomp}
\usepackage[hidelinks,bookmarks=false,draft]{hyperref}

\usepackage{url}
\usepackage{array,ragged2e}
\newcolumntype{P}[1]{>{\RaggedRight\arraybackslash}p{#1}}

\usepackage[font=small,skip=2pt]{caption}
\usepackage{textcomp, gensymb}
\usepackage{dblfloatfix}

\usepackage{mwe} 
\usepackage{microtype}

\usepackage[hidelinks, bookmarks=false]{hyperref}

\begin{document}
\title{Multimodal Optical Imaging Platform for Quantitative Burn Assessment}

\name{%
\begin{tabular}{c}
Nathaniel Hanson$^{\star}$\thanks{Nathaniel Hanson and Mateusz Wolak contributed equally.}, 
Mateusz Wolak$^{\star}$, 
Jonathan Richardson$^{\star}$\\[3pt]
\textit{Patrick Walker$^{\dagger}$, 
David M. Burmeister$^{\dagger}$, 
and Chakameh Jafari$^{\star}$\thanks{Correspondence: chakameh.jafari@ll.mit.edu.}}
\end{tabular}
}

\address{$^{\star}$Massachusetts Institute of Technology, Lincoln Laboratory, Lexington, MA, USA\\
         $^{\dagger}$Uniformed Services University, School of Medicine, Bethesda, MD, USA}

\ninept
\maketitle 

\begin{abstract}
Accurate assessment of burn severity at injury onset remains a major clinical challenge due to the lack of objective methods for detecting subsurface tissue damage. This limitation is critical in battlefield and mass-casualty settings, where rapid and reliable evaluation of burn depth is essential for triage and surgical decision-making.
We present a multimodal optical imaging framework that establishes the foundation for a compact, low-size, weight, and power (low-SWaP) field-deployable device for quantitative burn assessment. The system integrates broadband hyperspectral imaging (VSWIR, 400–2100 nm) with laser speckle contrast imaging to jointly evaluate biochemical composition and microvascular perfusion. Using short-wave infrared (SWIR, >1000 nm) wavelengths, we developed and validated novel deep-tissue parameters linked to water, lipid, and collagen absorption features that enhance burn-tissue separability and burn severity classification.
We implemented and validated unsupervised learning methods for spectral feature extraction, band down-selection, and clustering against histology, establishing a foundation for a rugged, data-driven device for early quantitative burn evaluation in austere environments.

\end{abstract}

\section{Introduction}
\label{sec:intro}
\vspace{-0.75em}

Accurate assessment of burn severity during the acute phase, 96 hours post-injury, remains a major clinical challenge~\cite{rehou2019acute}. Early classification of tissue viability is critical for resuscitation, debridement, and surgical planning. Within this window, burn wounds undergo complex biochemical and microvascular changes, including protein denaturation, collagen disruption, edema formation, and progressive ischemia that are not apparent from surface appearance alone~\cite{rogers2016managing}.

Recent advances in optical imaging have improved the noninvasive evaluation of burn depth; however, no single modality provides comprehensive biochemical and perfusion information. Ultrasound techniques can estimate dermal thickness and edema but remain operator-dependent and lack molecular specificity, reducing diagnostic consistency~\cite{ref4}. Laser Speckle Contrast Imaging (LSCI) offers rapid perfusion mapping but has limited depth sensitivity~\cite{ref1}. Spatial Frequency Domain Imaging (SFDI) quantifies absorption and scattering related to hemoglobin and water content but requires sequential illumination, restricting real-time use and making it susceptible to motion artifacts~\cite{ref2}. Hyperspectral Imaging (HSI) provides label-free biochemical contrast~\cite{schulz2022burn}; however, many of the existing studies measure only reflectance in the visible-near infrared (VNIR, 400 -- 1000 nm) \cite{chin2015hyperspectral, schulz2022burn, promny2022first}. Extending into the shortwave infrared (SWIR, $>$ 1000 nm) improves sensitivity to water, lipid, and collagen markers \cite{wilson2015review}. Finally, other systems use more multi-spectral systems with limited bands that target known spectral absorbance features \cite{thatcher2016multispectral, li2015outlier}.

\begin{figure}[!t]
    \centering
    \includegraphics[width=\linewidth]{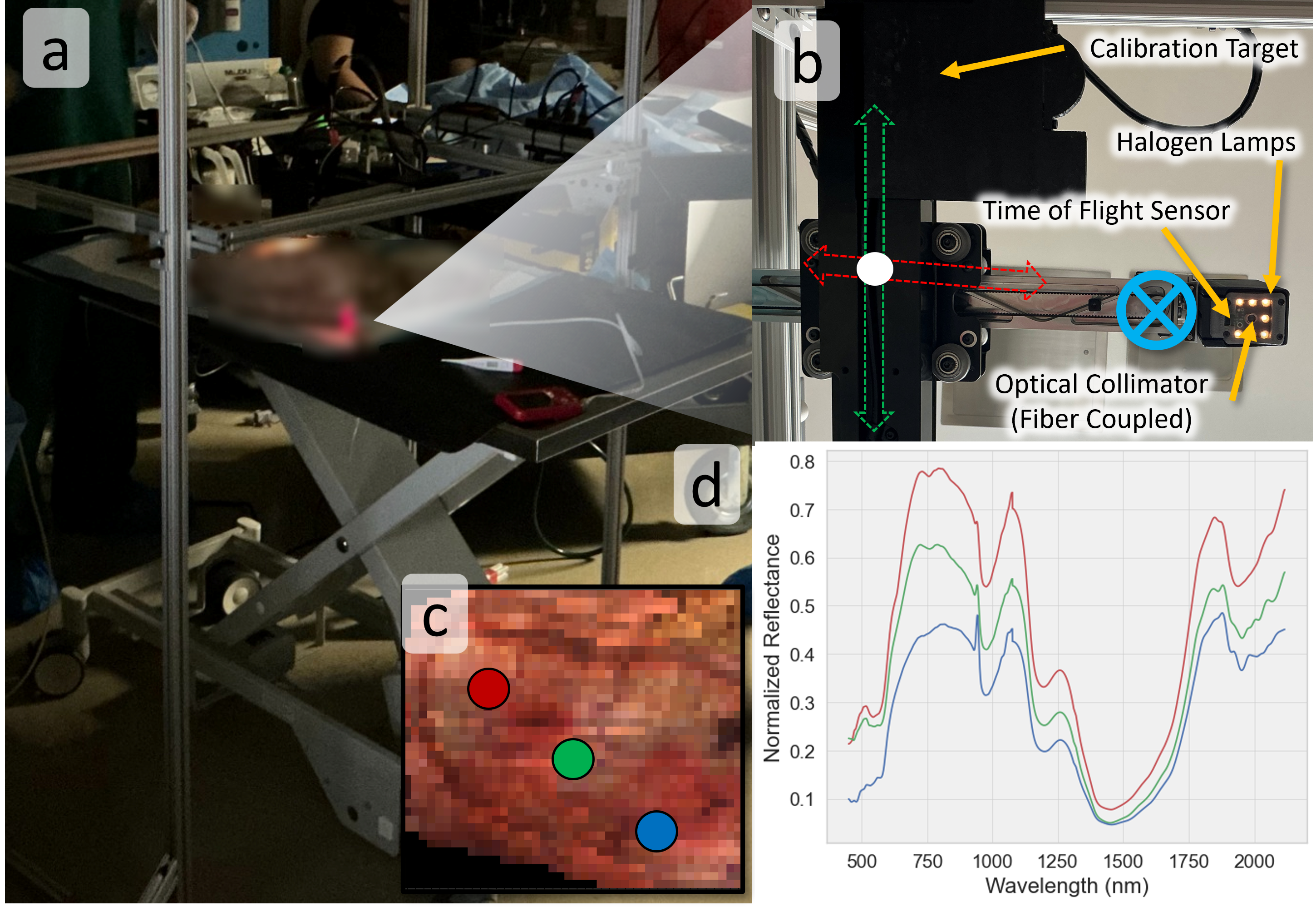}
    \caption{\textbf{Scanning Spectrometer Robot for Burn Imaging.} (a) Operating theater with porcine specimen. (b) Multi-modal sensor pod with axes of motion ($z$-axis into page) viewed from bottom of gantry. (c) RGB colorized hyperspectral image of burn region. (d) Extracted spectrum from areas of three regions of tissue in the sample from (c).}
    \label{fig:teaser}
    \vspace{-1.5em}
\end{figure}
While this range of imaging modalities has been explored, their sensitivity and specificity are limited during the early post-injury phase, when perfusion and inflammatory dynamics evolve rapidly. These challenges are exacerbated in conflict zones where thermobaric and incendiary weapons produce complex burn and blast injuries that require rapid, objective triage tools. 

To address these gaps, we developed a multimodal, robotic optical system (Fig.~\ref{fig:teaser}) that combines point spectroscopy on a rastering platform to iteratively build high spectral resolution hyperspectral datacubes (VSWIR, 400–2100 nm). Our system exceeds the spectral range and resolution of any previous research or commercially available system. Collected datacubes are spatially registered with Laser Speckle Contrast Imaging (LSCI) for label-free, quantitative assessment of burn injury. We register the imaging modalities and apply recent advances in unsupervised learning to perform automated spectral feature extraction and tissue segmentation. 

This paper contributes the following to the state-of-the-art:
\begin{itemize}
    \item Robotic hyperspectral imaging platform for iterative datacube acquisition across the visible-to-shortwave infrared.
    \item Introduction of novel deep-tissue physiological metrics derived from shortwave infrared bands, validated through multimodal correlation with perfusion and histological ground truth.
    \item Implementation of an unsupervised concrete autoencoder framework for feature extraction and wavelength down-selection.
\end{itemize}

\vspace{-0.75em}
\section{Data Collection}
\vspace{-0.75em}
\subsection{Porcine Model and Experimental Details}
Burn imaging was conducted in collaboration with military clinical partners at the Uniformed Services University (USU) using an established animal burn model. Sexually mature male Sinclair minipigs (Sinclair Bio Resources, Auxvasse, MO) were acclimated for at least 72~h prior to procedures. Animals were anesthetized with intramuscular tiletamine–zolazepam (Telazol, 4–8~mg/kg) and maintained under isoflurane (1–3\%) in 100\% oxygen following intubation and mechanical ventilation. Analgesia was provided with buprenorphine-HCl (0.01–0.05~mg/kg IM).

Full-thickness thermal burns covering approximately 25\% total body surface area (TBSA) were created using brass contact blocks (5$\times$5~cm and 9$\times$15~cm) heated to 100~$\pm$~0.2$^\circ$C for 30~s. Anatomical curvature introduced natural heterogeneity in burn depth, producing regions of mixed partial- and full-thickness injury.

Multimodal imaging, including hyperspectral imaging (HSI) and laser speckle contrast imaging (LSCI), was performed on two separate subjects: one immediately post-injury (day~0) and another seven days post-injury to evaluate temporal changes in perfusion and biochemical composition. All procedures were approved by the Institutional Animal Care and Use Committee (IACUC).
\subsection{Spectral Imaging}

While previous systems have utilized HSI for burn assessment in initial clinical studies, these devices have limited spectral resolution and range. Towards a broader understanding of burn severity, we utilize a point-based spectrometer to measure individual spatial points at high spectral resolution. This present device builds on our previous work in modeling spectra to create a sparse spectral representation of complex surfaces \cite{hanson2024prospect}. A coupled fiber optic cable transmits light from the surface into a spectrometer's entrance slit. As the fiber gathers light from a small discrete spatial area, we utilize a linear $x-y$ gantry system (Bantam Tools) to translate the perpendicular fiber over the surface of the tissue sample.

The fiber is contained within a custom sensor pod to enable consistent spectral sample acquisition. The fiber optic cable is coupled to an uncoated VSWIR collimator (Thorlabs). Surface illumination is provided by miniature incandescent NDIR lamps (Labsphere) through an uncoated borosilicate plate glass diffuser. As our specimens are measured while sedated, we must contend with thoracic excursion; this presents the added challenge of a variable observed spot size and illumination intensity. To compensate, the sensor pod has a miniature time-of-flight (ToF) distance sensor that measures the distance from the optical aperture to the specimen at 10 Hz. The pod is mounted to a linear rail, allowing $\pm 2$ cm of adjustment on the $z$ axis.

With knowledge of the absolute positioning of the gantry and its encoder position, we associate an $x,y$ position with every spectral measurement. To form the hyperspectral datacube, a desired rectangular region, with a defined width and height, is decomposed into a series of voxel positions with a small amount of spatial overlap. The gantry moves from position to position, pausing at each location for the spectrometer's integration time to allow the device to collect a continuous measurement of a single surface point. The list of observed spectra is then reshaped into a 3D hyperspectral datacube.

The robot measurement architecture is intentionally agnostic regarding the choice of spectrometer. Our initial HSI prototype employed an ultra-high-resolution VSWIR system (ASD FieldSpec 4 Hi-Res NG , Malvern Panalytical) to capture spectral information of the target tissue from 400 to 2500 nm. Our second study replaced the FieldSpec with two spectrometers --- VNIR (BlueWave, StellarNet, 350 -- 1100 nm) and SWIR (Rock XNIR, Ibsen Photonics, 1100 -- 2100 nm) --- and a bifurcated fiber optic cable. Fig.~\ref{fig:teaser}c shows the output of this iterative scanning and exemplar spectral measurements in Fig.~\ref{fig:teaser}d. For simplicity and to ensure consistency across spectrometer setups, we only consider the range of mutual spectral information in our further modeling (400–2100 nm).

The gantry contains a dark and white reference calibration target. The device auto-calibrates by collecting a dark measurement without active or ambient illumination and a measurement of a white reference target (99\% Spectralon, Labsphere) before each scanning process. The two references convert the device digital counts to reflectance using the formulation from \cite{hanson2024prospect}.

\subsection{Laser Speckle Contrast Imaging}
The Laser Speckle Contrast Imaging (LSCI) subsystem provides real-time maps of microvascular perfusion by analyzing speckle contrast dynamics generated under coherent laser illumination ($\lambda = 785$~nm) with camera-based detection. When coherent light illuminates the tissue surface, a random granular speckle pattern forms and fluctuates due to the motion of scatterers, primarily red blood cells. The rate and magnitude of these fluctuations reflect blood flow velocity, allowing estimation of relative perfusion.

LSCI data were captured using a Moor FLPI-2 device (Moor Instruments). Perfusion indices were computed from short-exposure image sequences using temporal speckle contrast analysis, where the contrast $K_t$ is defined as the ratio of the temporal standard deviation $\sigma_t$ to the mean intensity $\langle I_t \rangle$ across $N$ frames. Regions with higher flow exhibit faster fluctuations and lower contrast, while static or poorly perfused tissue yields higher contrast. Both temporal and spatial speckle contrast processing were explored and optimized to balance dynamic range and sensitivity, with temporal averaging improving robustness to noise and spatial filtering enhancing detection of localized perfusion changes.
\vspace{-0.75em}
\section{Modeling}
\vspace{-0.75em}
\subsection{Spatial and Spectral Preprocessing}
Hyperspectral reflectance datacubes were preprocessed for spectral consistency and multimodal alignment. HSI data were spatially aligned to the corresponding white-light images using fiducial markers and manual registration. LSCI images were automatically registered in Fiji ~\cite{Schindelin2012Fiji} using built-in alignment tools, followed by manual refinement to ensure accurate correspondence with the HSI field of view. Background masking removed non-tissue regions and fiducials, while spectral cropping excluded noisy or redundant bands. Wavelength-dependent smoothing corrected illumination non-uniformity and sensor noise, producing spatially and spectrally calibrated datacubes.

Additional preprocessing included per-pixel $\mathcal{L}_2$ normalization to mitigate illumination variability and per-band $Z$-score normalization to balance spectral intensity across wavelengths, ensuring uniform weighting in downstream analyses.

\subsection{Spectral Analysis and Unsupervised Segmentation}
\subsubsection{Dimensionality Reduction and Clustering}
\begin{figure}[!t]
    \centering
    \includegraphics[width=\linewidth]{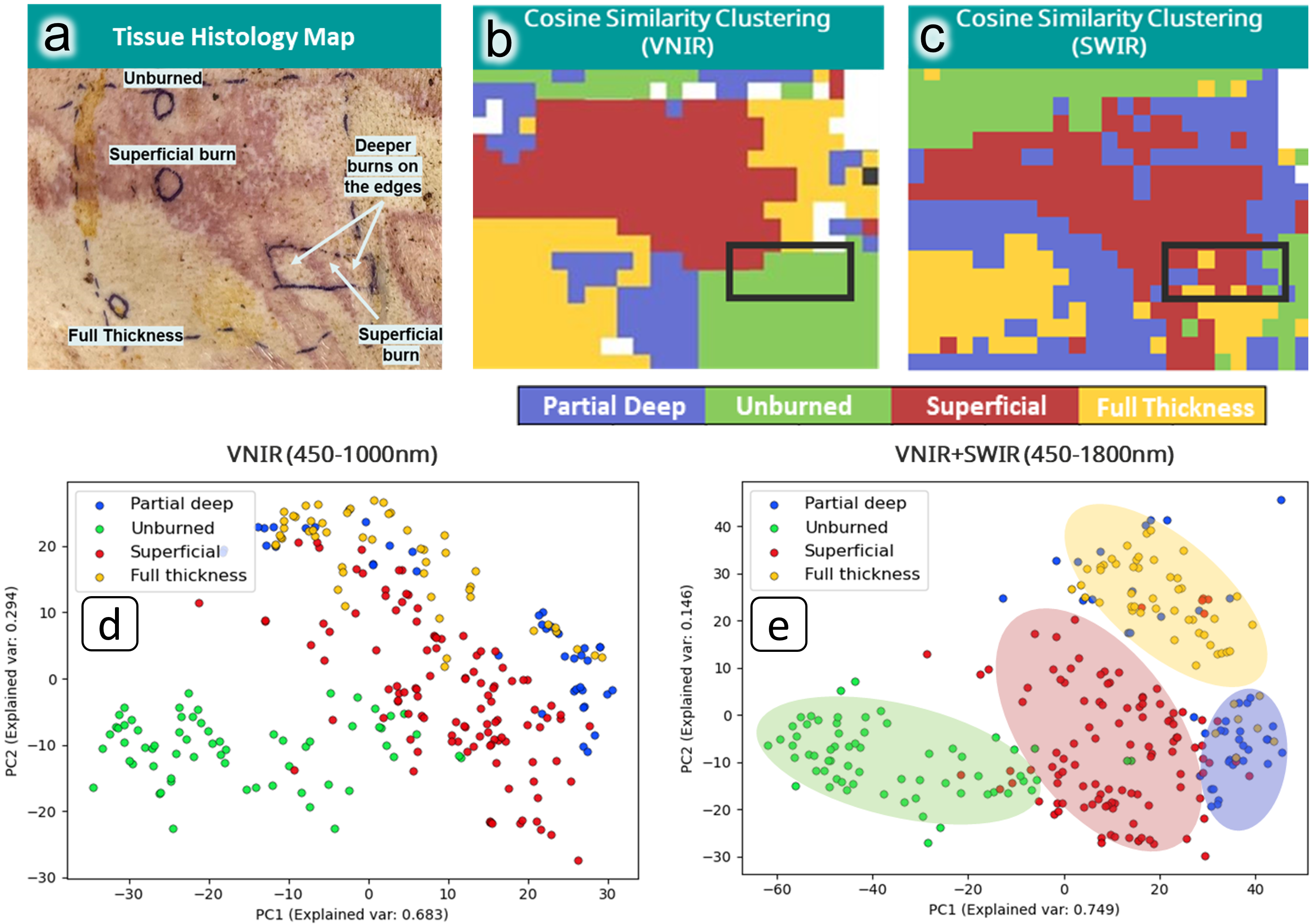}
    \caption{\textbf{Comparison of Tissue Segmentation and Spectral Separability Between VNIR and SWIR Domains}. (a) Ground-truth burn-severity histology labels. (b) VNIR clustering (450–1000 nm) fails to resolve tissue heterogeneity within the black-boxed region. (c) SWIR clustering accurately delineates superficial and deep/full-thickness burns within the highlighted regions. (d) PCA variance explained for VNIR bands indicates limited separability. (e) PCA variance explained for combined VNIR+SWIR bands demonstrates enhanced cluster separation and spectral discrimination. }
    \label{fig:swir}
    \vspace{-1.5em}
\end{figure}
Following preprocessing, unsupervised learning methods were applied to extract and cluster spectral features across the HSI dataset. Dimensionality reduction was performed using Principal Component Analysis (PCA) to capture variance associated with biochemical and structural alterations, followed by t-Distributed Stochastic Neighbor Embedding (t-SNE). Subsequent clustering using Gaussian Mixture Models (GMM), cosine-similarity-based spectral grouping, and k-means enabled delineation of tissue regions corresponding to superficial, partial-thickness, and deep burn injuries without reliance on extensive histological ground-truth.
\subsubsection{Tissue Physiological Feature Extraction}
Physiological parameters—including tissue oxygen saturation (StO$_2$)~\cite{ref5}, Deep Tissue Water Index (DTWI), and other deep-tissue metrics—were derived from hyperspectral absorbance data to capture biochemical variations across burn depths. Short-wave infrared (SWIR) bands were analyzed using spectral derivatives to enhance sensitivity to absorption features of water, collagen, lipid, and hemoglobin. The DTWI, defined as
\begin{equation}
\mathrm{DTWI} = \frac{\dfrac{A_{[1150-1230\,\mathrm{nm}]}}{A_{[1250-1350\,\mathrm{nm}]}} - s_2}{s_1 - s_2},
\end{equation}
is the normalized ratio between water absorption bands corresponding to an inflection point in the water absorption curve. The parameters $s_1$ and $s_2$ were empirically chosen heuristics to improve robustness under varying illumination and scattering conditions. These spectral indices provide label-free estimates of tissue oxygenation and structure, enhancing discrimination of burn severity and tissue viability.

\subsubsection{Concrete Autoencoder-based Clustering}
\label{sec:cac}
In addition to previous unsupervised analyses of the porcine burn dataset, a Concrete Autoencoder (CAE)~\cite{ref6} was implemented to perform differentiable feature selection and dimensionality reduction on hyperspectral data from the second study. Given the limited availability of histologically labeled ground truth, this approach enabled the extraction of informative spectral features and identification of latent clusters associated with tissue viability and burn depth. The CAE combines a nonlinear encoder–decoder architecture with a differentiable feature-selection layer that uses a Concrete (Gumbel–Softmax) relaxation defined as:
\begin{equation}
z_i = \frac{\exp((\log \alpha_i + g_i)/T)}{\sum_j \exp((\log \alpha_j + g_j)/T)},
\end{equation}
where $\alpha_i$ denotes the learned importance weight for each spectral band, $g_i$ represents stochastic noise that promotes exploration, and $T$ is a temperature parameter controlling the sharpness of selection. This mechanism selects a discrete subset of wavelengths that preserves key spectral information. The resulting compact embedding improves unsupervised burn-severity clustering by emphasizing physiologically relevant features and reducing redundancy.

Separate CAE models were trained for the StellarNet (1,502 bands) and Ibsen (255 bands) data to account for differing spectral resolutions. Each model employed a concrete selector layer to reduce inputs to five key bands, followed by a decoder composed of a dense layer with five units and a reconstruction layer matching the input size, both using LeakyReLU activations. $Z$-score–normalized cubes were flattened, split into training and validation sets, and trained with MSE loss and the Adam optimizer~\cite{kingma2017adammethodstochasticoptimization} (lr = 0.001), with the selector temperature exponentially decaying from 10 to 0.1 over 150 epochs.

At inference, randomly reconstructed spectra from the validation set were compared against their originals, and each scan was reconstructed using the full autoencoder. Outputs from both autoencoders were concatenated to restore the full spectral range. The 10 selected bands were retained for clustering. Hyperspectral reflectance data were $\mathcal{L}_2$- and $Z$-score-normalized, down-sampled to the selected bands, and compared via cosine-similarity matrices. These served as affinity matrices for spectral clustering, which embedded the affinities and applied $k$-means ($k = 4$) to generate burn severity labels.
\begin{figure}[!t]
    \centering
    \includegraphics[width=\linewidth]{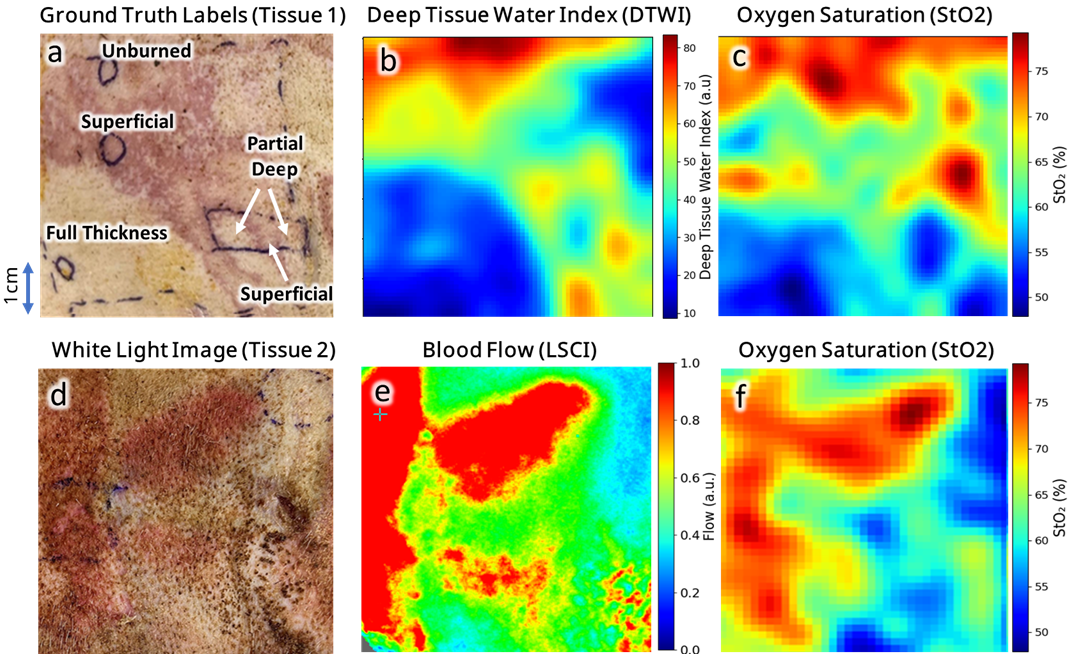}
    \caption{\textbf{Multimodal Correlation of Perfusion and Biochemical Parameters in Burn Tissue Using VNIR–SWIR HSI and LSCI} (a) Ground-truth burn-severity labels. (b) HSI-derived Deep Tissue Water Index (DTWI) map collected on day 0 of injury for Tissue 1 shows high index values in unburned areas (red) and low index in full-thickness regions (blue) consistent with collagen denaturation in deep injuries. (c) The corresponding StO$_2$ map for Tissue 1 highlights reduced perfusion in these same regions and correlates with DTWI trends. (d) White-light image of Tissue 2 (Day 7 post injury). (e) LSCI-derived blood-flow map (red: high flow, blue: low flow) and (f) HSI-derived StO$_2$ map show strong spatial agreement, validating the multimodal assessment of perfusion and biochemical state. Areas of high flow on Day 7 post injury correlated with inflammatory response in deep and full thickness burn after 3 days post injury}
    \label{fig:multimodal}
\end{figure}

\begin{figure}[!t]
    \centering
    \includegraphics[width=1.0\linewidth]{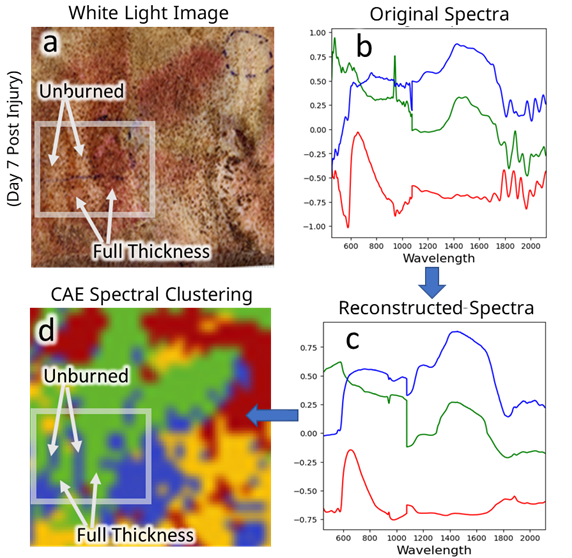}
    \caption{\textbf{Unsupervised CAE-Based Spectral Down-Selection and Clustering.} 
    (a) White-light image captured on day 7 post-injury with corresponding ground-truth histology labels shown in the boxed region. 
    (b) Original Z-scored and normalized spectra. 
    (c) Spectra reconstructed by the encoder using only 10 selected bands ($<1$\% of the original spectral dimensionality). 
    (d) Cosine-similarity clustering performed on the 10-band CAE down-sampled spectra. 
    Arrows indicate the sequential order of operations.}
    \label{fig:cae}
    \vspace{-2.5em}
\end{figure}
\vspace{-0.75em}
\section{Results}
\label{sec:results}
\vspace{-0.75em}
\subsection{Spectral Characterization and SWIR Sensitivity}
As shown in Figure~\ref{fig:swir}, inclusion of SWIR wavelengths ($>$1000~nm) markedly improved contrast between burn severities and enhanced spectral separability across tissue types. SWIR bands captured absorption peaks that are sensitive to water and collagen content, enabling better differentiation of viable and necrotic tissue compared to VNIR features alone.

Spectral clustering results demonstrated that VNIR-only data failed to resolve tissue heterogeneity within partial- and full-thickness burn regions (Fig.~\ref{fig:swir}b). Incorporating SWIR bands (Fig.~\ref{fig:swir}c) improved segmentation accuracy and clearly delineated superficial and deep burns consistent with histological ground truth (Fig.~\ref{fig:swir}a). Principal component analysis further confirmed this improvement, with combined VNIR+SWIR data (Fig.~\ref{fig:swir}d--e) yielding more distinct spectral cluster separation across burn severities.

\subsection{Multimodal Perfusion and Spectral Correlation}
Perfusion maps derived from temporal speckle contrast analysis revealed spatial gradients corresponding to areas of reduced microvascular flow, which aligned closely with HSI-derived parameters such as StO$_2$ and DTWI. As shown in Fig.~\ref{fig:multimodal}, superficial burns exhibited regions of high perfusion (low speckle contrast) that coincided with elevated StO$_2$ and DTWI values, reflecting preserved vascular integrity and tissue hydration. Conversely, full-thickness burns showed markedly reduced perfusion and diminished optical absorption near 970 and 1200~nm, consistent with dehydration and collagen denaturation in deep injuries.

At Day~7 post-injury, LSCI-derived perfusion maps and HSI-derived StO$_2$ distributions exhibited strong spatial correspondence (Figs.~\ref{fig:multimodal}d--f). Areas of increased flow correlated with inflammatory response in deep and full-thickness burns, confirming the complementary sensitivity of LSCI and VNIR--SWIR HSI to vascular and biochemical changes. Together, these results establish a robust multimodal framework for quantitative assessment of perfusion and tissue viability following thermal injury.

\subsection{CAE-based Feature Extraction and Clustering Analysis }

Concrete autoencoders (CAE) provide a data-driven deep learning method for selecting the 10 most representative spectral bands from the aggregated data of the second study. The autoencoders for the StellarNet and Ibsen spectrometer bands achieved a validation loss of 0.0138 and 0.0084, respectively, at the end of training. After training, the autoencoders were used for inference as described in Section~\ref{sec:cac} on random validation pixels. With < 1\% of the original bands, the autoencoder effectively recreates the general features of the original spectra, as seen in (b) and (c) in Fig.~\ref{fig:cae}. 

The final combined bands selected by the autoencoders were 528, 617, 762, 837, 954, 1,119, 1,324, 1,567, 1,775, and 2,005 nm. These bands were used to downsample the spectra from the second scan in the study, resulting in a 10-band hypercube. Next, cosine similarity spectral clustering assigned a label to each pixel in this hypercube. (d) in Fig.~\ref{fig:cae} shows a colorized and smoothed visualization of these labels. Treating this as a segmentation, we can see that this approach does well in differentiating the regions in our histology report (a) and corresponds with our other derived biochemical parameters in Fig.~\ref{fig:multimodal}.

\section{Discussion}
\vspace{-0.75em}
The prototype system was developed and tested, demonstrating accurate tissue perfusion mapping and burn-severity classification with potential for field deployment. Integration of VNIR–SWIR hyperspectral imaging with LSCI enabled a complementary assessment of structural, vascular and metabolic features essential for accurate burn-depth assessment, underscoring the value of a multimodal approach. CAE–based band selection achieved substantial spectral compression without loss of diagnostic fidelity, supporting the development of a compact multimodal device. In particular, the selected features included multiple SWIR wavelengths, demonstrating improved performance over prior literature that relied solely on VNIR bands. Future work will expand labeled dataset with increased depth resolution and enhance algorithms to better model fused multimodal interactions for more robust segmentation and burn-depth classification.

\vspace{-1em}
\section{Conclusion}
\label{sec:conclusion}
\vspace{-0.75em}
This work introduces a multimodal VSWIR–LSCI framework that integrates broadband hyperspectral and perfusion imaging with spectral feature extraction and physiological feature mapping, providing a quantitative, label-free approach for the early, objective assessment of burn depth and tissue viability.

Preliminary results from porcine studies demonstrate that the inclusion of SWIR wavelengths significantly enhances the separability of burn severity clustering and classification, improving the accuracy of label-free, unsupervised analysis. These findings establish a strong foundation for the continued refinement of the AI-driven VSWIR–LSCI framework and future development of a compact, Low-SWaP, field-deployable device for real-time assessment of tissue viability in both clinical and operational environments.

\clearpage
\section{Compliance with Ethical Standards}
\label{sec:ethics}
An Association for Assessment and Accreditation of Laboratory Animal Care International (AAALACI) accredited facility was used to perform the study. The study was approved by the Institutional Animal Care and Use Committee (IACUC) of the Uniformed Services University of the Health Sciences with all procedures in accordance with standard protocols and regulations. Additionally, our study complied with Animal Research: Reporting In Vivo Experiments (ARRIVE) guidelines.

\section{Acknowledgments}
DISTRIBUTION STATEMENT A. Approved for public release. Distribution is unlimited.
This material is based upon work supported by the Department of the Air Force under Air Force Contract No. FA8702-15-D-0001 or FA8702-25-D-B002.  Any opinions, findings, conclusions or recommendations expressed in this material are those of the author(s) and do not reflect the official policy or position of the Uniformed Services University of the Health Sciences or the Department of Defense. © 2025 Massachusetts Institute of Technology.
Delivered to the U.S. Government with Unlimited Rights, as defined in DFARS Part 252.227-7013 or 7014 (Feb 2014). Notwithstanding any copyright notice, U.S. Government rights in this work are defined by DFARS 252.227-7013 or DFARS 252.227-7014 as detailed above. Use of this work other than as specifically authorized by the U.S. Government may violate any copyrights that exist in this work. The authors have no relevant financial or non-financial interests to disclose.
\bibliographystyle{IEEEbib}
\bibliography{refs}

@article{ref1,
  author    = {A. Dijkstra and others},
  title     = {Laser speckle contrast imaging, an alternative to laser Doppler imaging in clinical practice of burn wound care: derivation of a color code},
  journal   = {Burns},
  year      = {2023},
  volume    = {49},
  number    = {8},
  pages     = {1907--1915},
  month     = dec,
  doi       = {10.1016/j.burns.2023.04.009},
}

@article{ref2,
  author    = {R. Rowland and A. Ponticorvo and M. Baldado and G. T. Kennedy and D. M. Burmeister and R. J. Christy and N. P. Bernal and A. J. Durkin},
  title     = {Burn wound classification model using spatial frequency-domain imaging and machine learning},
  journal   = {Journal of Biomedical Optics},
  year      = {2019},
  volume    = {24},
  number    = {5},
  pages     = {1--9},
  month     = may,
  doi       = {10.1117/1.JBO.24.5.056007},
  pmid      = {31134769},
  pmcid     = {PMC6536007}
}

@article{promny2022first,
  title={First preliminary clinical experiences using hyperspectral imaging for burn depth assessment of hand burns},
  author={Promny, Dominik and Aich, Juliane and Billner, Moritz and Reichert, Bert},
  journal={Journal of Burn Care \& Research},
  volume={43},
  number={1},
  pages={219--224},
  year={2022},
  publisher={Oxford University Press US}
}

@article{thatcher2016multispectral,
  title={Multispectral and photoplethysmography optical imaging techniques identify important tissue characteristics in an animal model of tangential burn excision},
  author={Thatcher, Jeffrey E and Li, Weizhi and Rodriguez-Vaqueiro, Yolanda and Squiers, John J and Mo, Weirong and Lu, Yang and Plant, Kevin D and Sellke, Eric and King, Darlene R and Fan, Wensheng and others},
  journal={Journal of Burn Care \& Research},
  volume={37},
  number={1},
  pages={38--52},
  year={2016},
  publisher={Oxford University Press}
}

@article{li2015outlier,
  title={Outlier detection and removal improves accuracy of machine learning approach to multispectral burn diagnostic imaging},
  author={Li, Weizhi and Mo, Weirong and Zhang, Xu and Squiers, John J and Lu, Yang and Sellke, Eric W and Fan, Wensheng and DiMaio, J Michael and Thatcher, Jeffrey E},
  journal={Journal of biomedical optics},
  volume={20},
  number={12},
  pages={121305--121305},
  year={2015},
  publisher={Society of Photo-Optical Instrumentation Engineers}
}

@INPROCEEDINGS{hanson2024prospect,
  author={Hanson, Nathaniel and Lvov, Gary and Rautela, Vedant and Hibbard, Samuel and Holand, Ethan and DiMarzio, Charles and Padır, Taşkın},
  booktitle={2024 IEEE/RSJ International Conference on Intelligent Robots and Systems (IROS)}, 
  title={PROSPECT: Precision Robot Spectroscopy Exploration and Characterization Tool}, 
  year={2024},
  volume={},
  number={},
  pages={5244-5251},
  doi={10.1109/IROS58592.2024.10802210}}

@article{rehou2019acute,
  title={Acute phase response in critically ill elderly burn patients},
  author={Rehou, Sarah and Shahrokhi, Shahriar and Thai, Joanne and Stanojcic, Mile and Jeschke, Marc G},
  journal={Critical care medicine},
  volume={47},
  number={2},
  pages={201--209},
  year={2019},
  publisher={LWW}
}

@article{ref4,
  author    = {S. Lee and Rahul and H. Ye and others},
  title     = {Real-time burn classification using ultrasound imaging},
  journal   = {Scientific Reports},
  year      = {2020},
  volume    = {10},
  pages     = {5829},
  doi       = {10.1038/s41598-020-62674-9},
  url       = {https://doi.org/10.1038/s41598-020-62674-9}
}

@article{ref5,
  author    = {A. Studier-Fischer and others},
  title     = {Spectral characterization of intraoperative renal perfusion using hyperspectral imaging and artificial intelligence},
  journal   = {Scientific Reports},
  year      = {2024},
  volume    = {14},
  pages     = {17262},
  doi       = {10.1038/s41598-024-68280-3},

}

@inproceedings{ref6,
  author    = {Abid, Abubakar and Balin, Melis F. and Zou, James},
  title     = {Concrete Autoencoders for Differentiable Feature Selection and Reconstruction},
  booktitle = {Proceedings of the 36th International Conference on Machine Learning (ICML)},
  year      = {2019},
  pages     = {444--453}
}

@misc{kingma2017adammethodstochasticoptimization,
      title={Adam: A Method for Stochastic Optimization}, 
      author={Diederik P. Kingma and Jimmy Ba},
      year={2017},
      eprint={1412.6980},
      archivePrefix={arXiv},
      primaryClass={cs.LG},
      url={https://arxiv.org/abs/1412.6980}, 
}

@article{wilson2015review,
  title={Review of short-wave infrared spectroscopy and imaging methods for biological tissue characterization},
  author={Wilson, Robert H and Nadeau, Kyle P and Jaworski, Frank B and Tromberg, Bruce J and Durkin, Anthony J},
  journal={Journal of biomedical optics},
  volume={20},
  number={3},
  pages={030901--030901},
  year={2015},
  publisher={Society of Photo-Optical Instrumentation Engineers}
}

@article{chin2015hyperspectral,
  title={Hyperspectral imaging for burn depth assessment in an animal model},
  author={Chin, Michael S and Babchenko, Oksana and Lujan-Hernandez, Jorge and Nobel, Lisa and Ignotz, Ronald and Lalikos, Janice F},
  journal={Plastic and Reconstructive Surgery--Global Open},
  volume={3},
  number={12},
  pages={e591},
  year={2015},
  publisher={LWW}
}

@article{schulz2022burn,
  title={Burn depth assessment using hyperspectral imaging in a prospective single center study},
  author={Schulz, Torsten and Marotz, J{\"o}rg and Seider, Sebastian and Langer, Stefan and Leuschner, Sebastian and Siemers, Frank},
  journal={Burns},
  volume={48},
  number={5},
  pages={1112--1119},
  year={2022},
  publisher={Elsevier}
}

@article{rogers2016managing,
  title={Managing severe burn injuries: Challenges and solutions in complex and chronic wound care},
  author={Rogers, Alan D and Jeschke, Marc G},
  journal={Chronic Wound Care Management and Research},
  pages={59--71},
  year={2016},
  publisher={Taylor \& Francis}
}

@article{Schindelin2012Fiji,
  title        = {Fiji: an open-source platform for biological-image analysis},
  author       = {Schindelin, Johannes and Arganda-Carreras, Ignacio and Frise, Erwin and Kaynig, Verena and Longair, Mark and Pietzsch, Tobias and Preibisch, Stephan and Rueden, Curtis and Saalfeld, Stephan and Schmid, Benjamin and Tinevez, Jean-Yves and White, Daniel J. and Hartenstein, Volker and Eliceiri, Kevin and Tomancak, Pavel and Cardona, Albert},
  journal      = {Nature Methods},
  volume       = {9},
  number       = {7},
  pages        = {676--682},
  year         = {2012},
  doi          = {10.1038/nmeth.2019},
  publisher    = {Nature Publishing Group},
  url          = {https://doi.org/10.1038/nmeth.2019}
}
\end{document}